\documentclass[twocolumn,twocolappendix,trackchanges]{aastex701}

\usepackage{amsmath}
\usepackage{gensymb}
\shorttitle{Dust Polarization Super-Resolution}
\submitjournal{ApJ}
\graphicspath{{./}}

\begin{document}

\title{Super-resolving Polarized Dust Emission with Transformer-Based Multi-Tracer Fusion}

\author[0000-0002-2679-4609]{Minjie Lei}
\affiliation{Department of Physics, Stanford University, Stanford, CA 94305, USA}
\affiliation{Kavli Institute for Particle Astrophysics \& Cosmology, P.O. Box 2450, Stanford University, Stanford, CA 94305, USA}
\email{minjilei@stanford.edu}

\author[0000-0003-2221-3018]{George Halal} 
\affiliation{Department of Physics, Stanford University, Stanford, CA 94305, USA}
\affiliation{Kavli Institute for Particle Astrophysics \& Cosmology, P.O. Box 2450, Stanford University, Stanford, CA 94305, USA}
\email{georgehalal@alumni.stanford.edu}

\author[0000-0002-7633-3376]{S. E. Clark}
\affiliation{Department of Physics, Stanford University, Stanford, CA 94305, USA}
\affiliation{Kavli Institute for Particle Astrophysics \& Cosmology, P.O. Box 2450, Stanford University, Stanford, CA 94305, USA}
\email{susanclark@stanford.edu}

\author{Alejandro Dobles}
\affiliation{Department of Computer Science, Stanford University, Stanford, CA 94305, USA}
\email{alejandro@genesistherapeutics.ai}

\author{Katie Brown}
\affiliation{Department of Physics, Stanford University, Stanford, CA 94305, USA}
\affiliation{Kavli Institute for Particle Astrophysics \& Cosmology, P.O. Box 2450, Stanford University, Stanford, CA 94305, USA}
\email{katienb@stanford.edu}

\author[0000-0002-7113-0262]{Viraj Manwadkar}
\affiliation{Department of Physics, Stanford University, Stanford, CA 94305, USA}
\affiliation{Kavli Institute for Particle Astrophysics \& Cosmology, P.O. Box 2450, Stanford University, Stanford, CA 94305, USA}
\email{virajvm@stanford.edu}

\begin{abstract}
Realistic models of polarized Galactic dust emission at high angular resolution are essential for component separation in next-generation cosmic microwave background (CMB) $B$-mode experiments. However, existing dust polarization templates are constrained only to ${\sim}1\degree$, and it remains challenging to develop models that can generate realistic small-scale complexity that is also coherently connected to the well-measured large-scale structures. In this work, we develop a data-driven super-resolution framework that predicts small-scale dust polarization structure directly from multiple tracers observed at high resolution. We combine the Planck dust optical depth $\tau_{353}$, \ion{H}{1}-based Stokes templates, and coarse Planck GNILC $Q$ and $U$ maps in a transformer-based model trained to recover native-resolution GNILC polarization from a $4\times$ beam-smoothed version of itself, so that the network solves a deconvolution problem informed by the structure of multiple tracers rather than generating structure to match a target statistic. Applying the model to the uniform-resolution GNILC maps, we produce $Q$ and $U$ predictions at $20\arcmin$ over the high-latitude ($|b|>30\degree$) sky. Using scalar and tensorial Minkowski functionals and scattering transform statistics, we find that our maps exhibit more coherent, anisotropic, and filamentary small-scale structure than the latest PySM dust model. Our results establish data-driven learned deconvolution as a promising route to foreground models whose small scales are inferred from data rather than prescribed.
\end{abstract}


\section{Introduction} \label{sec:intro}

The cosmic microwave background (CMB) is a foundational probe of the early universe, encoding both its initial conditions and its subsequent evolution \citep{hu02-cm}. While measurements of temperature anisotropies have yielded the bulk of existing cosmological constraints, current efforts are increasingly directed toward CMB polarization, and in particular the $B$-mode component at large angular scales \citep{kamionkowski16-bm}. Primordial $B$ modes are a direct signature of a stochastic gravitational wave background, a key prediction of inflationary theories \citep{seljak97-cmb, kamionkowski97-cmb}, and their detection is the primary science target of a broad program of ground-based \citep{cmbs4-16-sb, bk21-im, polarbear22-mb, spt23-mb, act24-dr, simons26-new}, balloon-borne \citep{spider22-cm}, and satellite \citep{litebird23-pt} experiments.

There are several observational challenges obscuring the pursuit of primordial $B$-mode detection. Gravitational lensing of CMB photons by intervening large-scale structure converts $E$ modes into $B$ modes that dominate the primordial signal at small angular scales \citep{planck20-lens}. While this lensing signal is itself cosmologically informative, accessing the underlying primordial component requires that it be reconstructed and removed to high precision \citep{hu02-ml}. The more formidable obstacle is Galactic foreground emission, primarily thermal dust and synchrotron, which dominate the polarized sky across frequencies \citep{dunkley09-fg, planck16-cs}.

Galactic foregrounds are moreover significantly non-Gaussian \citep{so19-fc}, a property that is expected to persist to small angular scales given the complex, turbulent structure of the interstellar medium (ISM) and the interstellar magnetic field. Non-Gaussian foregrounds couple modes in the angular power spectra and can bias both primordial $B$-mode inference and lensing reconstruction, though this can be mitigated with foreground cleaning \citep{beck20-bi}. Realistic simulations of the polarized dust sky at high angular resolution are therefore a prerequisite for validating component separation pipelines at the sensitivity of upcoming experiments. Such simulations are currently limited by the absence of high-resolution, large-area polarization observations. Existing observational constraints are derived largely from Planck and \textit{WMAP} data, and measure the polarized sky with high fidelity only down to approximately $1\degree$ \citep{planck20-fg, bennett13-wm}. Simulation frameworks such as the Python Sky Model \citep[PySM;][]{thorne17-pm, zonca21-pm} and its predecessor the Planck Sky Model \citep{delabrouille13-psm} extend these templates to smaller scales by fitting a power law to the observed polarization power spectra and populating the unconstrained scales with a Gaussian realization, while extrapolating across frequencies under assumptions about the dust spectral energy distribution. The most recent PySM dust models improve on this by generating small scales in a log polarization tensor space, so that the injected fluctuations become non-Gaussian when transformed back to Stokes parameters \citep{borrill25-pm}. The small scales nonetheless remain a stochastic realization conditioned on a global summary statistic, and are not tied to the morphology of the observed sky. Complementary approaches include data-driven templates built from independent tracers \citep{clark19-hs, bk23-de, halal24-hi}, phenomenological models \citep{hervias22-ph}, and magnetohydrodynamic (MHD) simulations \citep{kim19-mh}, each of which faces limitations in either reproducing the observed morphology or reaching high resolution efficiently.

Recent advances in machine learning offer a route to high-resolution, non-Gaussian foreground simulations that are learned from the data rather than prescribed. Generative adversarial networks have been used to inject small-scale structure into low-resolution dust observations while preserving their statistical properties, reaching $12\arcmin$ in ForSE \citep{krachmalnicoff21} and $3\arcmin$ in ForSE+ \citep{yao24forse}. However, these models are trained on high-resolution dust total intensity maps and calibrated against the Minkowski functionals of the intensity small scales, then applied to Stokes $Q$ and $U$ separately. They therefore assume that the small-scale statistical properties of polarized dust emission match those of total intensity. Moreover, by conditioning on global summary statistics, the predicted small-scales are not necessarily constrained to produce morphological features coherent with the large-scale structures.

In this work, we propose a data-driven super-resolution approach that complements and advances existing approaches in two important ways: by learning the small scales directly from polarization data, so that the model solves a deconvolution problem rather than generating structure to match a target statistic; and by conditioning the prediction on multiple high resolution tracers of polarized dust emission structure, including \ion{H}{1} emission. The physical basis for the latter is well established. Dust polarization is closely connected to the morphology of the neutral ISM, with magnetically aligned \ion{H}{1} filaments tracing the plane-of-sky field orientation \citep{clark15-hi, clark19-hs}, and the degree of alignment between filaments and the local field imprinting the observed $E$/$B$ asymmetry and $TE$ correlation \citep{cuckierman23-hs, halal24-hi}. Motivated by these results, we combine the Planck dust optical depth $\tau_{353}$ and \ion{H}{1}-based polarization templates with the coarse GNILC Stokes maps in a transformer-based model trained to recover native-resolution dust polarization from a smoothed version of itself.

The rest of the paper is organized as follows. In Section \ref{sec:data}, we describe the Planck and \ion{H}{1} datasets used in this study and the construction of the training set. Section \ref{sec:method} introduces the super-resolution model architecture, along with the Minkowski functional and scattering transform statistics used to validate it. In Section \ref{sec:results}, we present the training and validation results across input resolutions, and examine the attention maps of the trained model. Section \ref{sec:prediction} describes the application of the model to produce high-resolution polarization maps over the high-latitude sky. In Section \ref{sec:validation}, we quantify the non-Gaussianity of the predicted small scales and compare against Gaussian realizations and the latest PySM dust model. Finally, we discuss the implications and limitations of the method in Section \ref{sec:discussion}, followed by conclusions in Section \ref{sec:conclusion}.

\section{Data} \label{sec:data}

\subsection{Training Datasets} \label{subsec:train_data}

\subsubsection{Planck Dust Emission Maps at 353 GHz} \label{subsubsec:planck_pol}
The target of our super-resolution task is the dust polarized emission maps from Planck at 353 GHz. We use maps that separate the cosmic infrared background (CIB) radiation from the Galactic dust emission using the Generalized Needlet Internal Linear Combination technique \citep[GNILC,][]{remazeilles11-gnl, olivari16-gn}. We further add a Galactic offset of 63 $\mu K_{\rm CMB}$ and CIB monopole subtraction of 452 $\mu K_{\rm  CMB}$ to the GNILC total intensity map \citep{planck18-gn}. We utilize the variable-resolution version of the Planck R3.00 GNILC component-separated map to produce the Q and U image patches for training, and the uniform-resolution version for inference. The process of creating image patches from full sky maps are described in more detail in Section \ref{subsec:pre-process}.

Additionally, we utilize for model training the R1.20 Planck dust optical depth map at 353 GHz. The optical depth $\tau$ is obtained by fitting a modified blackbody spectrum to Planck total intensity data \citep{planck15-od}. This provides a high-resolution input map tracing dust spatial distribution at 5\arcmin\ angular resolution. 

\subsubsection{\ion{H}{1} Polarization Template} \label{subsubsec:hi_pol}
\ion{H}{1} and dust are strongly correlated over the diffuse regions of sky that are of particular interest to CMB observations \citep{lenz17-hd}.  Additionally, \ion{H}{1} emission is free from CIB contamination and contains no correlated instrumental systematics with dust emission observations. Utilizing the Rolling Hough Transform, \citet{clark14-ma} found that \ion{H}{1} intensity distribution exhibits filamentary structures that are significantly aligned with local magnetic field orientation. Building on these results, \citet{clark19-hs} developed a template based solely on \ion{H}{1} intensity and measures of \ion{H}{1} filament orientation, and showed that the \ion{H}{1}-based template is strongly correlated with Planck 353 GHz polarization data. 

We utilize the \ion{H}{1}-based Stokes Q and U templates as additional inputs to our model. The specific version of the \ion{H}{1} template we use is described in \citet{halal24-hi} and uses a Hessian-based filament finding algorithm originally described in \citet{cuckierman23-hs}. The algorithm is applied to the \ion{H}{1}4PI dataset, which is a full-sky \ion{H}{1} survey at 16.2\arcmin\ angular resolution and 1.49 km/s spectral resolution \citep{hi4pi16-rl}.  We integrate the Q and U templates along the velocity range -13 km/s $<$ $v$ $<$ 16 km/s, which \citet{halal24-hi} found results in the strongest correlation with Planck 353 GHz data. The final data products are converted to the COSMO convention to match the Planck convention. 

\begin{figure}[t]
    \centering
    \includegraphics[width=0.48\textwidth]{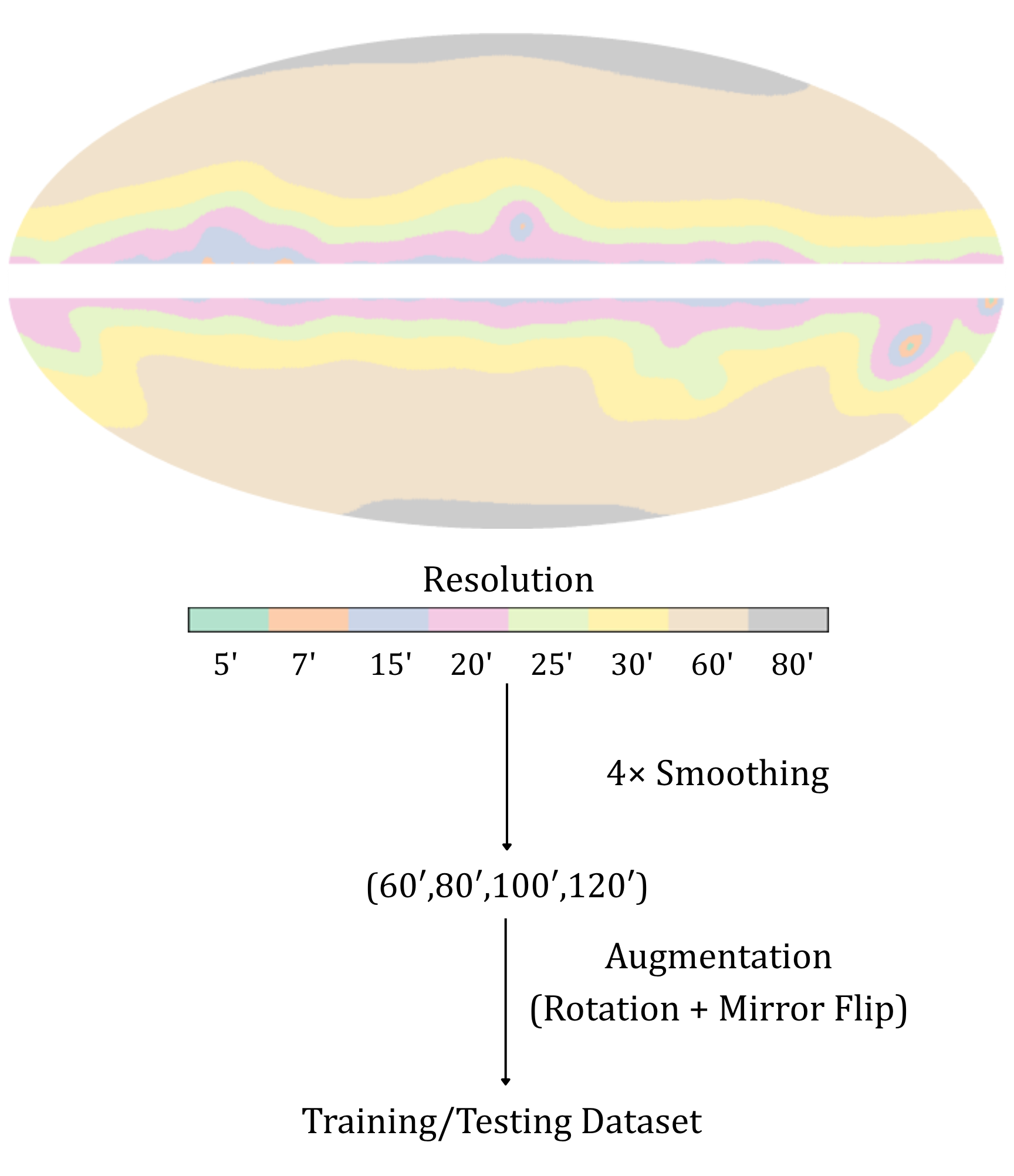}
    \caption{Schematic of the training dataset construction. The top map shows the color-coded resolution regions of the variable-resolution Planck GNILC map, ranging from $5\arcmin$ to $80\arcmin$. The $15\arcmin$, $20\arcmin$, $25\arcmin$, and $30\arcmin$ regions are smoothed by a factor of four to $60\arcmin$, $80\arcmin$, $100\arcmin$, and $120\arcmin$ to form the low-resolution inputs, and the resulting patches are augmented through rotation and mirror-flip augmentation to build the training and testing datasets.}
    \label{fig:model_training}
\end{figure}

\subsection{Data Preprocessing} \label{subsec:pre-process}
To create projected image patches for model training, we reference the different resolution regions used in the variable-resolution GNILC map shown in Figure \ref{fig:model_training}. For the ground truth images, we do not use simulations or reference existing templates. Instead, the highest-resolution Planck GNILC data available is used as the high-resolution ground truth, and the model is trained on artificially smoothed version of the ground truth to recover the higher-resolution input.

From the different resolution regions, we smooth the GNILC Stokes $Q$ and $U$ images at ($15\arcmin$, $20\arcmin$, $25\arcmin$, $30\arcmin$) to ($60\arcmin$, $80\arcmin$, $100\arcmin$, $120\arcmin$). As a result, the model is trained to up-scale the resolution of the input images by a factor of 4 regardless of the input resolution. The $5\arcmin$, $7\arcmin$ and $10\arcmin$ resolution regions are not included since the regions are too small to create training patches. The $60\arcmin$ and $80\arcmin$ regions are also not used since the $4\times$ smoothed images would be too low resolution. The resulting training patches cover $\sim 40\%$ of the sky. The final smoothed maps are normalized to span $[0,1]$ for $\tau_{353}$ and [-1, 1] for the Stokes maps. The normalization procedure is described in more detail in Appendix \ref{appx:hyperparameter}.

From the smoothed full sky maps, we create patch projections for training and testing. For each target resolution $\theta_{\rm target}\in($15\arcmin$, $20\arcmin$, $25\arcmin$, $30\arcmin$)$, we project patches of physical size $20\theta_{\rm target}\times20\theta_{\rm target}$, with $80\times80$ pixelization. The neighboring patches partially overlap such that the longitudinal and latitudinal distances between patch centers are $8\theta_{\rm target}$. At each $\theta_{\rm target}$, patch projections are created from all regions with a resolution higher than or equal to $\theta_{\rm target}$, smoothed to $\theta_{\rm target}$. To augment the dataset, we repeat the patch projection procedure for maps after applying $45\degree$, $90\degree$, $180\degree$, and $270\degree$ rotations. We also apply a mirror flip for each rotation. In total, the augmentation results in 29,464 image patches including the original images, for each of the five input channels  ($\tau_{353}$, $Q_{HI}$, $U_{HI}$, $Q_{353}^{LR}$, $U_{353}^{LR}$), as well as the ground truth ($Q_{353}^{HR}$, $U_{353}^{HR}$) before the $4\times$ artificial smoothing is applied. Finally, the image patches are split into 80\% training, 10\% validation, and 10\% testing datasets. 

\subsection{Validation Datasets} \label{subsec:val_data}

\subsubsection{Gaussian Realization} \label{subsubsec:gauss_real}
As a baseline to compare our high resolution dust polarization prediction against, we generate a Gaussian realization of dust polarization features at small-scales. The detailed procedure follows what is described in \citet{thorne17-pm}. The input Planck GNILC polarization maps are first smoothed with a Gaussian kernel of FWHM = $2.6\degree$ to filter out small-scale features. Polarization power spectra from the resulting spectra are fitted with a power law in the range of multipole $\ell=20-69$ and extended to larger multipoles. Finally, smaller scale features are generated from Gaussian realizations of the fitted power spectra. An amplitude modulated version is also produced by modulating the small scale Gaussian realization to match the amplitude of the original large-scale template following the same procedure outlined in \citep{borrill25-pm}. 

\subsubsection{Python Sky Model} \label{subsubsec:pysm}
In addition to Gaussian realization baselines, we also compare our prediction against current state-of-the-art foreground models.  We make use of  the Python Sky Model (PySM), a publicly available software that provides models of polarized dust and synchrotron emission at CMB frequencies constrained by the latest observational data \citep{zonca21-pm, borrill25-pm}. Specifically, we utilize the dust ``d10" model. To generate realizations of small-scale polarization features currently unconstrained by full-sky survey data, PySM introduces the ``polarization fraction tensor" formalism, where total intensity $I$ and polarized intensity $P$ are coupled by defining a symmetric rank-2 tensor: 
\begin{equation} \label{eqn:pol_ten}
    p_{ab} = \ln\begin{bmatrix} I+Q & U \\ U & I-Q \end{bmatrix} = \begin{bmatrix} i+q & u \\ u & i-q \end{bmatrix}
\end{equation}
where $i, q, u$ are the log polarization tensor analogues of the $I, Q, U$ stokes parameters. Defined this way, transformation to the emission maps are applied consistently as matrix operations to $p_{ab}$ , and the physical condition $I^2\geq Q^2+U^2$ is enforced by the criteria that $p_{ab}$ is positive semi-definite. The power law fitting process described in Section \ref{subsubsec:gauss_real} is then applied to the $i, q, u$ templates, with the resulting small-scale realizations modulated and added to the large scale maps before transforming back to $I, Q, U$ maps. Because the log polarization tensor transformation is non-linear, any Gaussian fluctuations introduced in $i, q$, and $u$ maps will necessarily become non-Gaussian when transformed back to I , Q, and U in a self-consistent way. 

\begin{figure*}[t]
    \centering
    \includegraphics[width=0.95\textwidth]{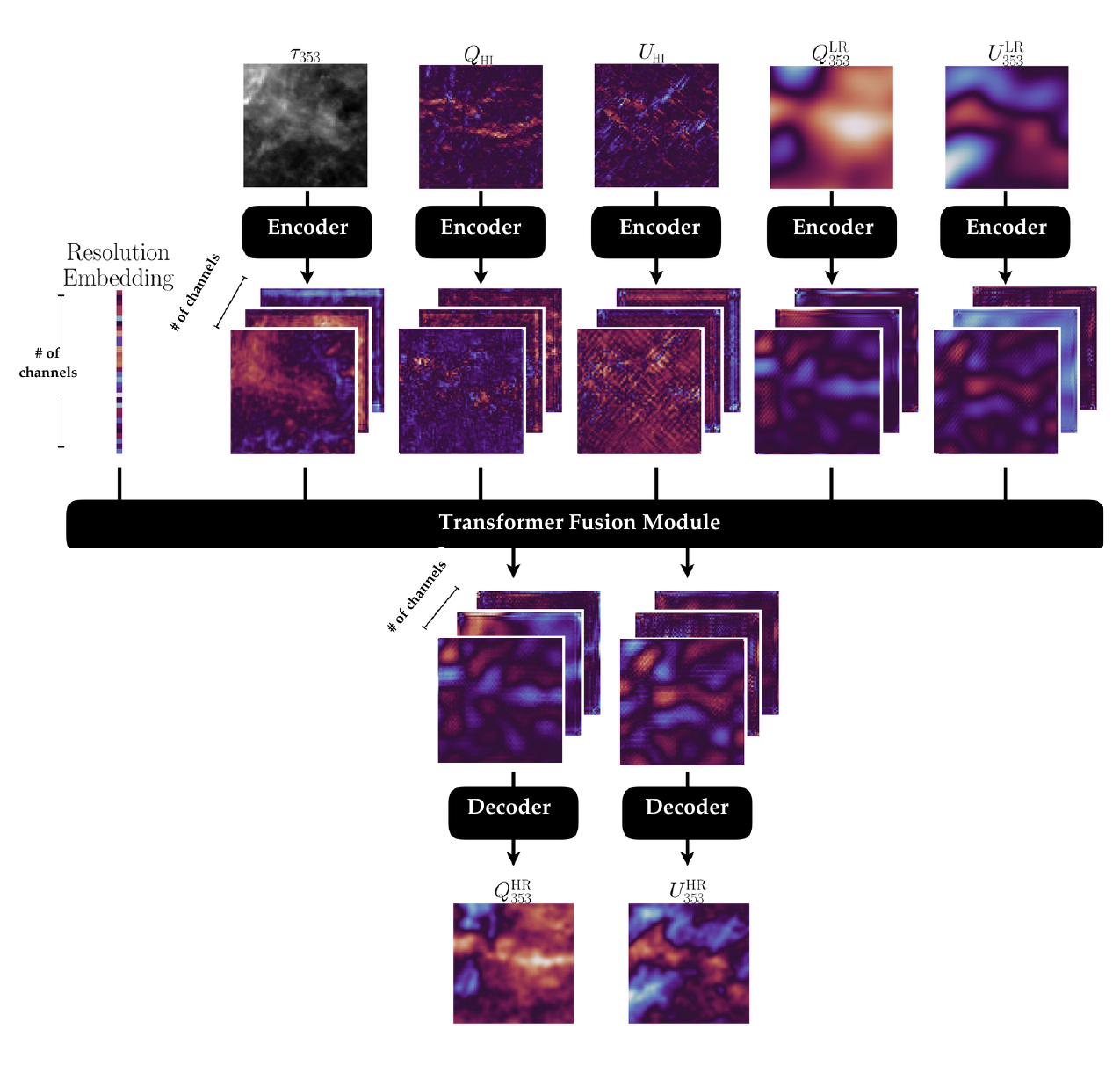}
    \caption{Schematic of the transformer-based foreground super-resolution network. Convolutional encoders map the input channels---Planck optical depth $\tau_{353}$, the \ion{H}{1}-based Stokes templates $Q_{HI}$ and $U_{HI}$, and the coarse GNILC $Q_{353}^{LR}$ and $U_{353}^{LR}$---into feature maps, while the target output resolution is encoded through a separate embedding layer. The transformer fusion module performs self-attention across all input channels, before being fed to a decoder to produce the final high-resolution predictions $Q_{353}^{HR}$ and $U_{353}^{HR}$.}
    \label{fig:model_architecture}
\end{figure*}

\section{Methodology} \label{sec:method}

\subsection{Problem Context} \label{subsec:problem}
In the context of computer vision, the task of generating and adding realistic small-scale features to coarse, large-scale images falls under the problem of image super-resolution. There are two typical approaches to this problem: single image super-resolution (SISR) and multi-image super-resolution (MISR). SISR techniques aim to enhance the resolution of a single coarse image, while MISR methods leverage multiple low-resolution images of the same scene to reconstruct a high-resolution output. Recently, deep learning-based MISR has transformed the field by introducing sophisticated mechanisms for fusing information from multiple low-resolution inputs \citep[e.g.][]{bordone20-ds}. More specifically, the use of transformers and the attention mechanism as the fusion module have seen a lot of successful applications \citep[e.g.][]{an22-tr, li23-mast}. The attention mechanism was originally introduced in natural language processing to allow models to learn different parts of the input as contexts in a non-sequential way. In the context of image super-resolution, the attention mechanism enables MISR models to selectively learn the importance of different spatial regions over different feature channels. This allows the model to be more expressive by capturing multi-scale information and cross-scale and cross-channel relationships. 

In this work, we propose a new approach to generating high-resolution polarized dust emission maps by fusing together dust and \ion{H}{1} information, utilizing transformer and the attention mechanism as the fusion module, building on the model first proposed in \citet{halal24-pd}.  As described in Section \ref{subsec:train_data}, we use both high-resolution dust optical depth map (5\arcmin) and \ion{H}{1} polarization template (16\arcmin) to inform high-resolution prediction from low-resolution polarized dust emission information. 


\subsection{Transformer-Based Super-Resolution Model} \label{subsec:model}
In Figure \ref{fig:model_architecture}, we show a schematic of the network architecture used for our transformer-based image super-resolution model. A single network is trained to predict both high-resolution stokes Q and U maps, taking advantage of the fact that the Q and U channels are not physically independent. To predict small-scale information from input low resolution $Q_{353}^{LR}$ and $U_{353}^{LR}$ maps, we utilize high-resolution tracers Planck optical depth $\tau_{353}$, and \ion{H}{1} filament-based Stokes $Q_{HI}$ and $U_{HI}$ maps. In addition to the high-resolution image tracers, we also encode the target output resolution as an input vector using an embedding layer. This will inform the scale-dependence of the model predictions.

To fuse together information from the multiple input tracers, we utilize a hybrid CNN+transformer architecture, with a transformer-based fusion module sandwiched by convolution-based encoders and decoders. The convolution module encodes the input images  ($\tau_{353}$, $Q_{HI}$, $U_{HI}$, $Q_{353}^{LR}$, $U_{353}^{LR}$) into multiple feature channels with $N_{\rm channels}=32$ without upsampling or downsampling, resulting in feature images with the same pixelation as the input images. The target output resolution is encoded into a vector of the same channel size $N_{\rm channels}=32$ through an embedding layer. The feature vectors are then fed through the transformer fusion module, with the self attention mechanism expressed as:
\begin{equation} \label{eq:attention}
    \text{Attention}(Q, K, V) = \text{softmax}\left(\frac{QK^T}{\sqrt{d_k}}\right)V
\end{equation}
where $Q$, $K$, $V$ are the learnable query, key, and value matrices, $V$ is the value matrix containing the input sequence, and $d_k$ is the attention head dimension, chosen to be 32 in this model.  After the final transformer layer, the first few channels involving target scale and ($\tau_{353}$, $Q_{HI}$, $U_{HI}$) are discarded, and the final two channels corresponding to ($Q_{353}^{LR}$, $U_{353}^{LR}$) are fed through a decoder to produce the final high resolution predictions ($Q_{353}^{HR}$, $U_{353}^{HR}$). More details about the model architecture can be found in \citet{halal24-pd} and Appendix \ref{appx:hyperparameter}. Hereafter we refer to this vision-transformer (ViT)-based approach as the dust ViT model. 

As described in Section \ref{subsec:pre-process},  each training sample to the model consists of 7 images plus a target resolution encoding. Five input image channels ($\tau_{353}$, $Q_{HI}$, $U_{HI}$, $Q_{353}^{LR}$, $U_{353}^{LR}$) , and the target resolution in arcmins, always at $4\times$ the resolution of the input $Q_{353}^{LR}$ and $U_{353}^{LR}$, are fed through the model to predict the final $Q_{353}^{HR}$, $U_{353}^{HR}$. L1 loss is used to compute the difference between the predicted and target high resolution images on a per channel/pixel basis. More details about the hyper-parameter choices and loss function used to train the model can be found in Appendix \ref{appx:hyperparameter} and Table \ref{tab:hyper}.

\subsection{Validation Statistics} \label{subsec:val_metric}

\subsubsection{Power Spectra} \label{subsubsec:cls}
Angular power spectra are commonly used to characterize temperature (TT) and polarization (EE and BB) variations across angular scales. We utilize the \texttt{NaMaster} software \citep{alonso19-nm} to compute $\mathcal{D}_{\ell}^{XX} \equiv \ell(\ell+1)C_{\ell}/2\pi$, where XX are TT, EE, and BB modes respectively. $C_\ell$ represents the pure-E and pure-B estimator, computed to eliminate E-B mixing on masked sky maps. We estimate the uncertainty of $D_{XX}$ assuming simple Gaussian sample variance and no instrumental noise using the following equation \citep{knox97-cm}: 
\begin{equation} \label{eq:sigma_ps}
    \sigma_{\mathcal{D}_{\ell}^{XY}} = \sqrt{\frac{\mathcal{D}_{\ell}^{XX} \mathcal{D}_{\ell}^{YY} + (\mathcal{D}_{\ell}^{XY})^{2}}{f_{\text{sky}} (2\ell + 1) \Delta \ell}}
\end{equation}
where $f_{\text{sky}}$ is the sky fraction of the mask used, and $\Delta\ell$ is the bin size chosen for the power spectra computation. 

\subsubsection{Scalar and Tensorial Minkowski Functionals} \label{subsubsec:mf_statistics}
To quantify the level of non-Gaussianity in our predicted high-resolution dust polarization maps, we make use of the scalar and tensorial Minkowski functionals \citep[MF,][]{minkowski1903}. These descriptors provide a complete morphological characterization of excursion sets—regions where the field intensity or polarization exceeds a specific threshold.  In the case of scalar MF computed on two-dimensional maps, we can define three descriptors $\mathcal{V}_0$, $\mathcal{V}_1$, $\mathcal{V}_2$, which corresponds to the area, perimeter and connectivity (Euler Statistic) respectively. By contrasting the behavior of $\mathcal{V}_0$, $\mathcal{V}_1$, $\mathcal{V}_2$ as a function of threshold $\rho$ to the expected behavior of a Gaussian map, we can quantify the deviation from Gaussianity of our maps. 

While scalar MFs are powerful morphological descriptors to characterize non-Gaussianity, they are inherently isotropic and insensitive to preferred orientations and structural elongations prevalent in the magnetized, turbulent ISM. To complement the scalar MF information, we utilize the tensorial extension of these statistics \citep{schroder13-mf}, which rather than counting scalar quantity features, map the distribution of orientations. Tensorial MFs can be constructed from integration of unit normal vectors $\hat{n}$:
\begin{align} \label{eq:tensorial_mf}
\mathcal{W}_1(\nu) &= \int \mathbf{n} \otimes \mathbf{n} \, dl \\
\mathcal{W}_2(\nu) &= \int \kappa \, (\mathbf{n} \otimes \mathbf{n}) \, dl
\end{align}
where $\kappa$ is the local curvature, and $dl$ is the line element integral over the excursion set. From the $\mathcal{W}_1$ and $\mathcal{W}_2$ tensors, we can solve for eigenvalues $\lambda_1$ and $\lambda_2$ and define \citep{schroder13-mf, joby19-tm}:
\begin{itemize}
    \item Alignment parameter: $\alpha=\lambda_2/\lambda_1$ where $\lambda_2\leq\lambda_1$. $\alpha$ measures the global anisotropy of the input field, with $\alpha=1$ indicating a globally isotropic field. Fields with global anisotropy have $\alpha<1$.
    \item Shape parameter: $\beta=\left<\lambda_{2,i}/\lambda_{1,i}\right>$, where $i$ denotes the individual connected components of the excursion set. $\beta$ describes the intrinsic anisotropy of average structures in the field. Highly elongated/filamentary fields have $\beta\sim0$, while $\beta=1$ corresponds to a field made up of perfect circles. Gaussian random fields have $\beta$ values around $0.7$.
    \item Orientation parameter: $\mathcal{O}=(1-\alpha)/(1-\beta)$ defines a normalized measure of collective structure alignment. In the case of total coherent alignment, the global anisotropic direction and the average individual structure direction align ($\alpha=\beta$), and $\mathcal{O}=1$.
\end{itemize}

\subsubsection{Scattering Transform Statistics} \label{subsubsec:st}
An additional set of statistics we utilize to quantify the non-Gaussianity of our predicted map is the scattering transform (ST). While the MFs provide a set of intuitive descriptors of the overall morphological features of the input maps, the ST statistics characterize morphological properties in a scale-dependent way. This is achieved by utilizing a set of oriented wavelets as bandpass filters in an iterative process to encode localized, non-Gaussian information from input maps at discrete scales and orientations \citep{mallat2012-gi, allys19-rw, cheng2021-si, cheng23-ss}. We utilize the Morlet family of wavelets for this study \citep{morlet82-wl}: 
\begin{equation}
    \psi_{j,l}(\boldsymbol{x})=2^{-2j}\psi_0(2^{-j}R_\theta^{-1}\boldsymbol{x})
\end{equation}
where $\psi_0$ is the mother wavelet defined as a sinusoidal wave localized by a Gaussian envelope.  $\psi_{j,l}$ is constructed from $\psi_0$ via a series of scaling and rotation operations indexed by $0\leq j\leq J$ and $\theta\in\{l\pi/L,0\leq l< L\}$. We choose L = 4 such that $\theta$ probes $(0\degree, 45\degree, 90\degree, 135\degree)$. The scales are sampled in powers of 2 and the maximum scale J is set by the dimension of the input map $N$ as $J=\log_2N-1$. Given input field $I(\boldsymbol{x})$, the ST operations are computed as the convolutions of wavelets with input fields followed by a pointwise modulus: $|I(\boldsymbol{x})\star\psi_{j,l}|$. Iterative application of the convolution and modulus operation gives rise to higher order coefficients. The first two orders of the ST are computed as:
\begin{eqnarray} \label{eq:st_eqn}
    S_1(j_1, l_1) =& \left<\left|I(\boldsymbol{x})*\psi_{j_1, l_1}(\boldsymbol{x})\right|\right> \\ \nonumber
    S_2(j_1, j_2, l_1, l_2) =& \left<\left|(\left|I(\boldsymbol{x})*\psi_{j_1, l_1}(\boldsymbol{x})\right|) * \psi_{j_2, l_2}(\boldsymbol{x})\right|\right> 
\end{eqnarray}
where $\left<\cdot\right>$ denotes spatial averaging to arrive at a set of summary statistics that is only a function of scales and orientations. The full $S_1$ and $S_2$ coefficients can be further reduced into combinations with more intuitive interpretations \citep{cheng20-an}. First, a feature quantifying ``sparsity" can be defined from the ratio of $S_2$ and $S_1$ coefficients averaged over orientations:
\begin{equation} \label{eq:s21}
    s_{21}=\left<S_2/S_1\right>_{l_1,l_2}
\end{equation}
As the field becomes more sparse and non-Gaussian, the strength of variation of the field moves increasingly  towards higher order coefficients. As a result, higher $s_{21}$ values indicate a stronger departure from Gaussian fields. Moreover, we can select specific orientation sets of the $S_2$ coefficient to define:
 \begin{eqnarray} \label{eq:st_po}
        S_{2\parallel}(j_1, j_2) = \left<S_2(j_1, j_2, |l_1-l_2|=0\degree, l_1)\right>_{l_1}\\
        S_{2\perp}(j_1, j_2) = \left<S_2(j_1, j_2, |l_1-l_2|=90\degree, l_1)\right>_{l_1}
    \end{eqnarray} 
where $S_{2\parallel}$ selects coefficients that probe aligned ($|l_1-l_2|=0\degree$) structures across different scales,  while $S_{2\perp}(j_1, j_2)$ probes anti-aligned ($|l_1-l_2|=90\degree$) structures. The ratio of $S_{2\parallel}$ and $S_{2\perp}$ defines a shape parameter that measures the filamentarity of the input fields: 
\begin{equation} \label{eq:s22}
    s_{22}= S_{2\parallel}/ S_{2\perp}
\end{equation}
For our analysis, we further condense the $s_{21}$ and $s_{22}$ coefficients by rearranging the terms as ($j_1$, $\Delta j=j_2-j_1$), before averaging over the smaller scales $j_1$. The resulting $s_{21}(\Delta j)$ and $s_{22}(\Delta j)$ are only functions of the relative scales $j_2-j_1$. A field dominated by thin filamentary structures will have large $s_{21}(\Delta j)$ and $s_{22}(\Delta j)$ values at large relative scales $\Delta j$. 

ST statistics have been successfully applied to a range of problems in astrophysics, including ISM phase separation \citep{lei23-st, lei25-st}, morphological characterization of diffuse Galactic light \citep{liu25-oc}, and the separation of cosmic infrared background \citep{blancard21-dn, delouis22-st, auclair24-st}. Here we utilize the reduced interpretable ST coefficients in conjunction with the scalar and tensorial MFs to evaluate the morphological features of our high-resolution map predictions in a comprehensive and scale-dependent way.

\begin{figure*}[t!]
    \centering
    \includegraphics[width=0.95\textwidth]{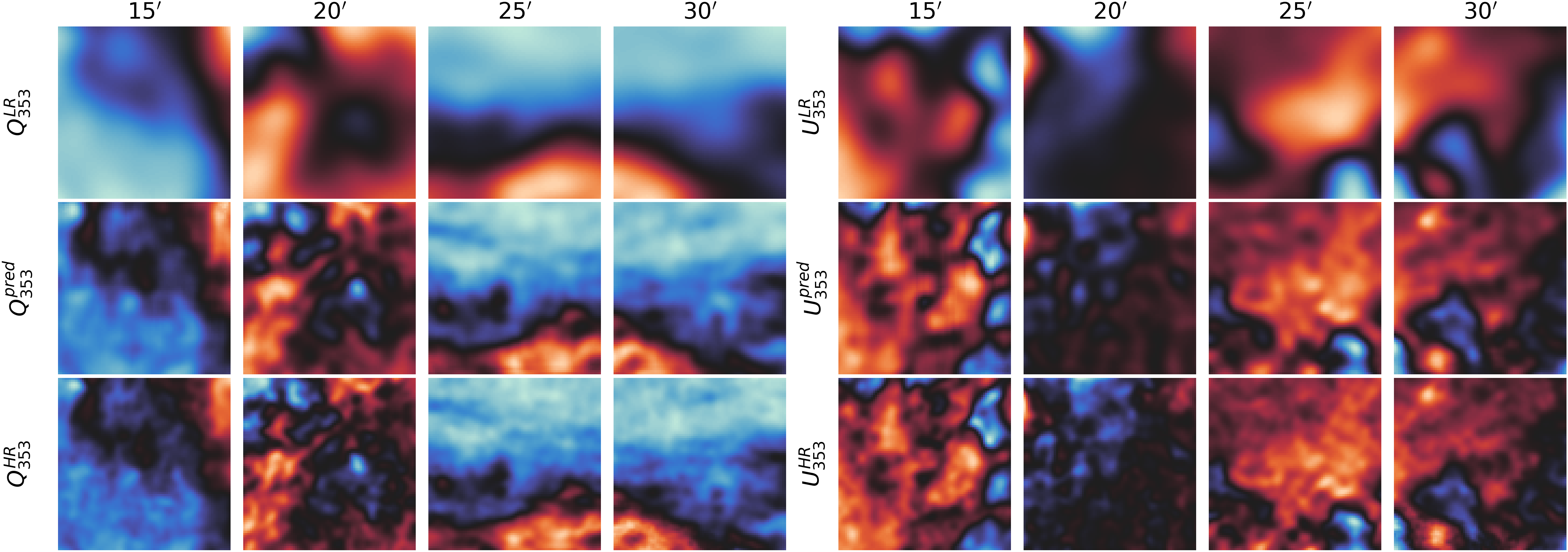}
    \caption{Super-resolution results for four validation patches, one per target resolution (columns: $15\arcmin$, $20\arcmin$, $25\arcmin$, $30\arcmin$). For each Stokes parameter we show the low-resolution input $Q_{353}^{LR}$ ($U_{353}^{LR}$) at $60\arcmin$, $80\arcmin$, $100\arcmin$, $120\arcmin$, the model prediction $Q_{353}^{pred}$ ($U_{353}^{pred}$), and the ground-truth $Q_{353}^{HR}$ ($U_{353}^{HR}$). The predictions agree closely with the ground truth across all resolutions, with a mean $L_1$ validation loss of $<0.003$ for maps normalized to $[-1, 1]$.}
    \label{fig:training_res}
\end{figure*}

\section{Super-Resolution Model Training Results} \label{sec:results}
In Figure \ref{fig:training_res}, we show the result of applying our super-resolution model to four sample images from the validation dataset at different resolutions the model is trained on. The input images $Q^{LR}_{353}$, $U^{LR}_{353}$ are at (60\arcmin, 80\arcmin, 100\arcmin, 120\arcmin) resolutions respectively, and we compare the predicted outputs $Q^{pred}_{353}$, $U^{pred}_{353}$ with the ground truth samples $Q^{HR}_{353}$, $U^{HR}_{353}$ at (15\arcmin, 20\arcmin, 25\arcmin, 30\arcmin). Over the validation dataset which the model did not see in training, the predicted images agree with the ground truth to a high extent, with an average $L_1$ validation loss of 0.003 for images normalized between -1 and 1. There is no significant difference in the model's performance over different resolutions.

\begin{figure}[t]
    \centering
    \includegraphics[width=0.45\textwidth]{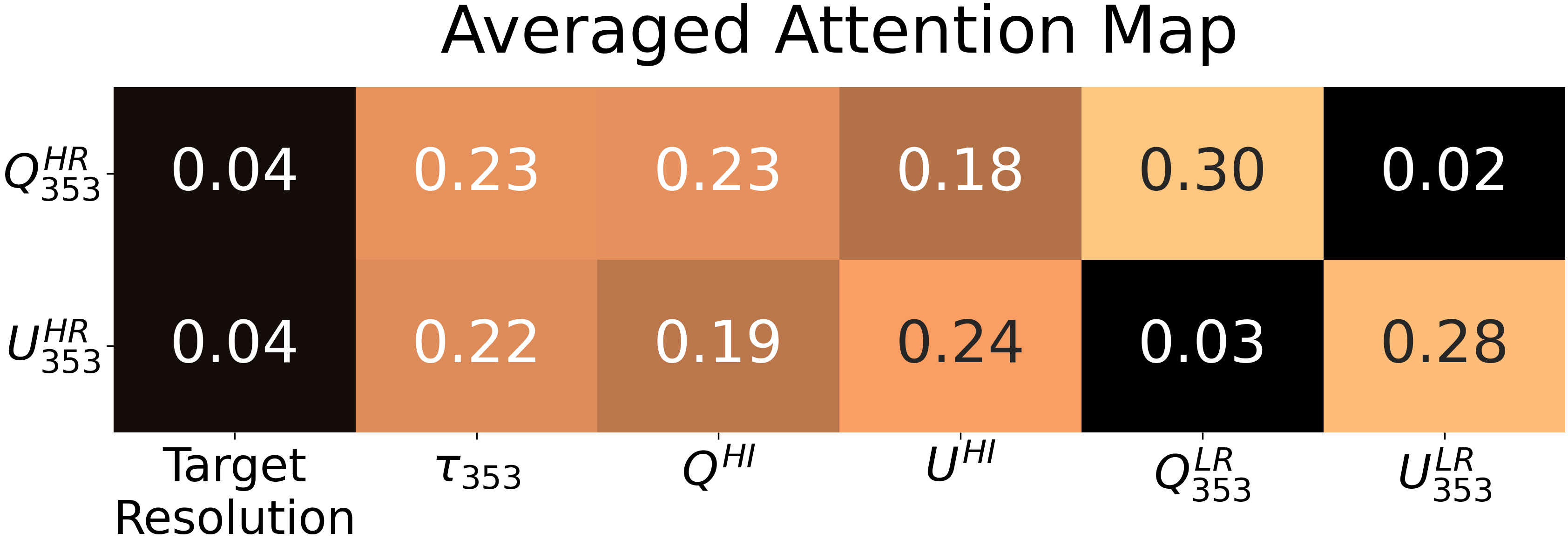}
    \caption{Attention weights of the trained model averaged over all pixels of 100 validation patches (25 per target resolution). Each row shows the relative importance of the six input channels---target resolution, $\tau_{353}$, $Q_{HI}$, $U_{HI}$, $Q_{353}^{LR}$, and $U_{353}^{LR}$---to predicting the high-resolution $Q_{353}^{HR}$ (top) and $U_{353}^{HR}$ (bottom) outputs, with the weights in each row summing to unity.}
    \label{fig:attention_averaged}
\end{figure}

\begin{figure*}[t]
    \centering
    \includegraphics[width=0.95\textwidth]{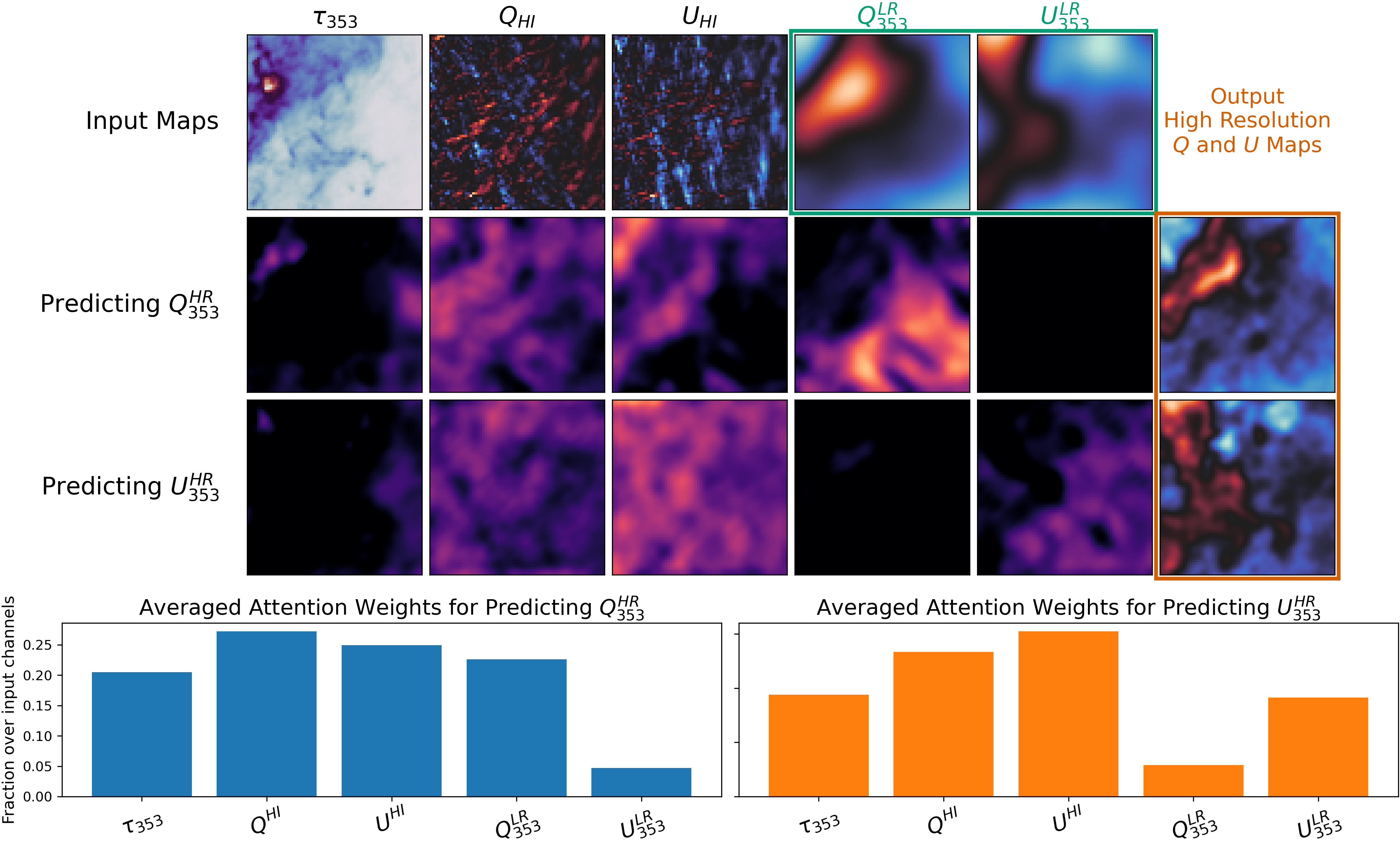}
    \caption{Per-channel attention maps for a sample test patch upscaled from $60\arcmin$ to $15\arcmin$. \emph{Top row}: the five input maps ($\tau_{353}$, $Q_{HI}$, $U_{HI}$, $Q_{353}^{LR}$, $U_{353}^{LR}$) together with the predicted high-resolution $Q$ and $U$ maps (right). \emph{Middle rows}: the spatial attention maps for the $Q_{353}^{HR}$ and $U_{353}^{HR}$ queries, each normalized to sum to one over input channels, for each query. \emph{Bottom row}: histograms of the per-channel attention fractions averaged over pixels for the two queries. The channels contribute complementary morphological information, with the \ion{H}{1}-based templates tracing the dense filamentary structure and $Q_{353}^{LR}$/$U_{353}^{LR}$ informing the diffuse background.}
    \label{fig:attention_image}
\end{figure*}

One advantage of the attention mechanism described by Equation \ref{eq:attention} is that the attention weights provide a direct way to interpret how important each input channel is to predicting the final high resolution maps. In Figure \ref{fig:attention_averaged}, we show the attention weights of the trained model averaged over all pixels of 100 patches, 25 patches each for the four different target resolutions. The averaged weights provide a way to compare the overall relevance of each type of input map ($\tau_{353}$, $Q_{HI}$, $U_{HI}$, $Q_{353}^{LR}$, $U_{353}^{LR}$) to predicting the final $Q_{353}^{pred}$ and $U_{353}^{pred}$ outputs. There are several key takeaways from Figure \ref{fig:attention_averaged}. First, as expected $Q^{LR}_{353}$ is important for predicting  $Q^{HR}_{353}$ while $U_{353}^{LR}$ is important for predicting $U^{HR}_{353}$. Interestingly, the high-resolution tracers dust optical depth $\tau_{353}$ and \ion{H}{1}-based Stokes $Q_{HI}$ and $U_{HI}$ are each comparably informative to $Q^{HR}_{353}$ and $U_{353}^{HR}$ predictions. $Q_{HI}$ ($U_{HI}$) are slightly more informative to predicting $Q^{HR}_{353}$ ($U^{HR}_{353}$) than $U^{HR}_{353}$ ($Q^{HR}_{353}$), as expected. Finally, as the first column of Figure \ref{fig:attention_averaged} shows, the target resolution information is not informative to predicting the high resolution output. This supports the picture that the turbulent-driven diffuse ISM appears self-similar across many spatial scales. The attention weight result together with the physical expectation of self-similarity supports the assumption that the dust foreground super-resolution task is partially scale-independent, and we can apply the model across different resolution regions of the input GNILC map in a similar way.

While the averaged attention map gives a sense of the overall relevance of each input channel to predicting high resolution outputs, we can also examine the full attention maps corresponding to each channel when applying the model to specific images. In Figure \ref{fig:attention_image}, we show the full attention maps from applying the dust ViT model to a sample test image upscaled from 60\arcmin\ to 15\arcmin\ resolution. The top panels show the input channel images and the per-channel attention maps.  Each row of the attention maps add up to one when summed over over input channels, for each query. In the bottom panel of Figure \ref{fig:attention_image} we show the histogram of attention weights per channel averaged over pixels. The histogram shows that different input channels (dust optical depth, $Q_{HI}$, $U_{HI}$, $Q_{353}^{LR}$ and $U_{353}^{LR}$) are comparably important to predicting high resolution dust polarization, but the full attention map reveals an even more interesting picture where each channel contributes in complementary ways in terms of morphological information to the final high resolution prediction. The sample input maps $Q_{353}^{LR}$ and  $U_{353}^{LR}$ both feature a prominent anisotropic dense structure over a more diffuse background. Looking at the attention maps querying $Q_{353}^{HR}$, the $Q_{353}^{LR}$ maps mostly informs the diffuse background regions, while $Q_{HI}$ and $U_{HI}$, the \ion{H}{1}-based maps constructed from filamentary structures, reveal dense anisotropic structure in their attention maps. $\tau_{353}$, the highest resolution input channel at 5\arcmin, appears to be more informative of smaller scale features. Together all the input channels combine to provide complementary information to the task of predicting high resolution polarized dust emission maps with realistic non-Gaussian features in a consistent and data-driven way.

\begin{figure*}[t]
    \centering
    \includegraphics[width=0.95\textwidth]{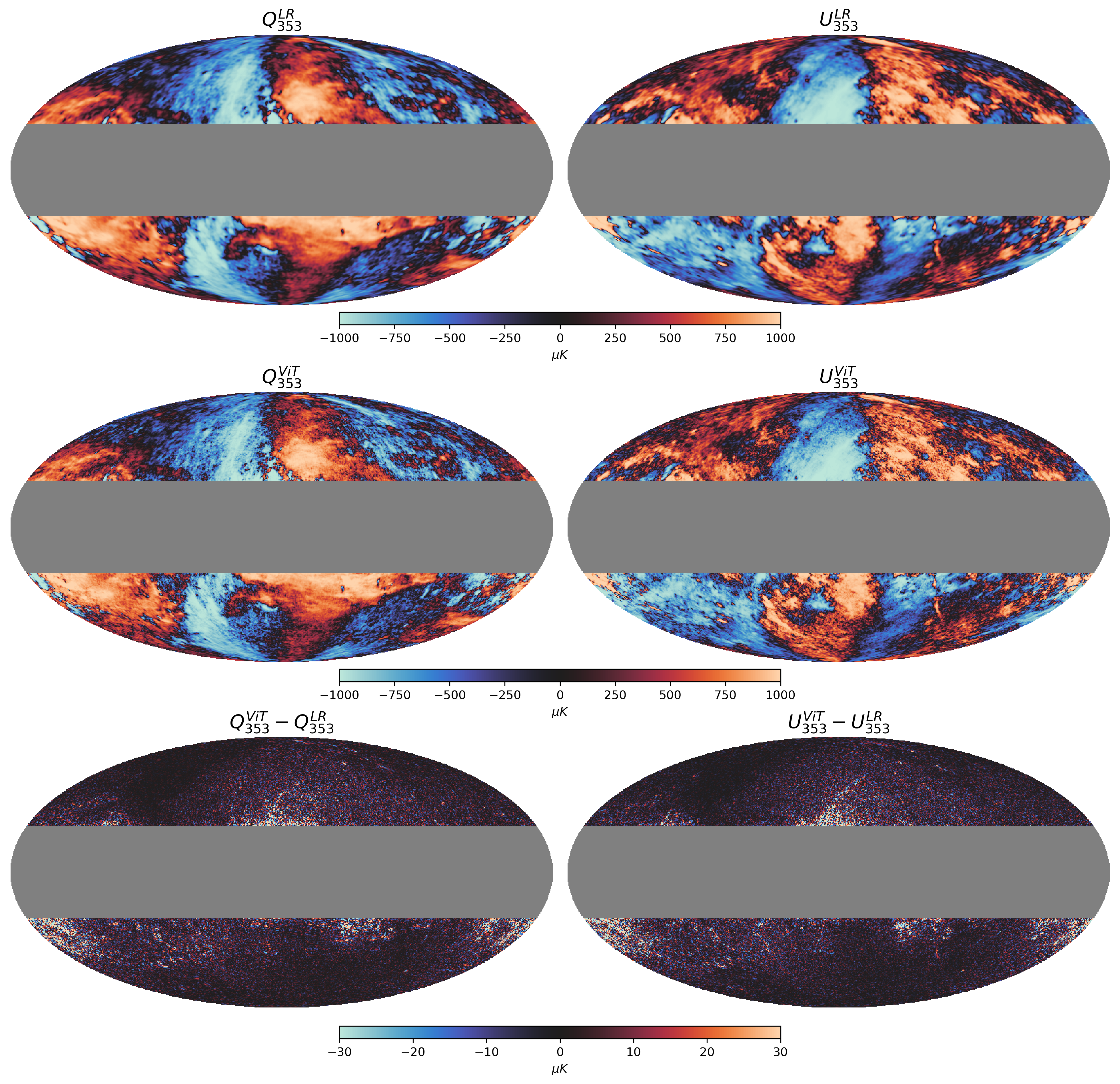}
    \caption{Full-sky, high-latitude ($|b|>30\degree$) super-resolution prediction from the dust ViT model. \emph{Top}: the uniform-resolution GNILC input maps $Q_{353}^{LR}$ and $U_{353}^{LR}$ at $80\arcmin$. \emph{Middle}: the Dust ViT model predictions $Q_{353}^{ViT}$ and $U_{353}^{ViT}$ at $20\arcmin$. \emph{Bottom}: the residual maps $Q_{353}^{ViT}-Q_{353}^{LR}$ and $U_{353}^{ViT}-U_{353}^{LR}$.}
    \label{fig:prediction_maps}
\end{figure*}

\begin{figure}[t]
    \centering
    \includegraphics[width=0.45\textwidth]{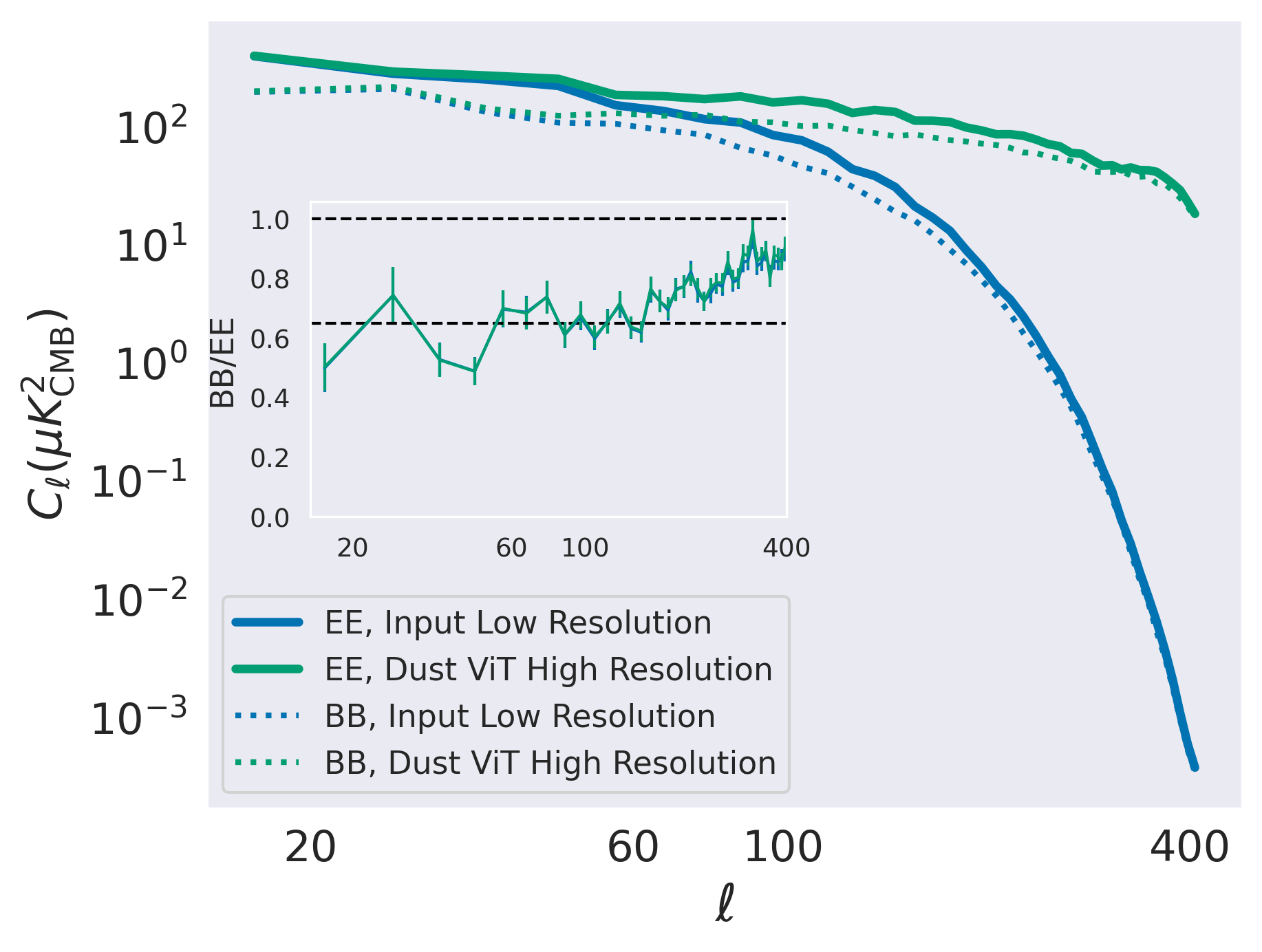}
    \caption{$EE$ and $BB$ angular power spectra $C_\ell$ of the input low-resolution GNILC maps (blue) and the Dust ViT high-resolution predictions (green). The inset shows the BB/EE ratio for both sets of maps. The high resolution prediction preserves the BB/EE ratio of the input GNILC map. }
    \label{fig:power_spectrum}
\end{figure}

\section{High-Resolution Dust Foreground Prediction} \label{sec:prediction}
Having validated the performance of the super-resolution model trained to $4\times$ the resolution of smoothed GNILC patches over various input resolutions at 60\arcmin, 80\arcmin, 100\arcmin\ and 120\arcmin, we apply the method to upscale raw, unsmoothed GNILC patches at high latitude regions not seen by the model during training. To produce uniform high-resolution, high latitude ($|b|>30\degree$) polarized dust emission maps, we utilize the uniform-resolution version of the GNILC map at 80\arcmin\ resolution as input. We follow the same procedure described in Section \ref{subsec:pre-process}. The sky is first projected into patches of physical size $20\theta_{\rm target}\times20\theta_{\rm target}$ with $80\times80$ pixelization, where $\theta_{\rm target}=80/4=20\arcmin$. The neighboring patches are overlapping by 50\% of the patch size. To avoid excessive overlapping towards higher latitudes, the longitudinal separation between patch centers are adjusted by a factor of $\cos{b}$. The super-resolution model is applied to each individual patch before reprojecting back to full sky in HEALPix format. When combining overlapping patches, a cosine window is applied so that central pixels are weighted higher than pixels closer to the boundary of a patch. 

In Figure \ref{fig:prediction_maps}, we show the predicted high latitude, high resolution $Q^{ViT}_{353}$ and $U^{ViT}_{353}$ maps at 20\arcmin\ resolution, alongside the input GNILC low resolution maps at 80\arcmin, and the residual maps computed as the difference between the high resolution prediction and the GNILC maps. The large-scale patterns of the predicted high-resolution map closely match that of the low resolution input, while the residual maps clearly encode small-scale structures as expected. 

Following the procedure described in Section \ref{subsubsec:cls}, we compute the $EE$ and $BB$ power spectra of the dust ViT map using a high latitude sky mask $|b|>30\degree$, and compare them with those of the input uniform resolution GNILC map at $80\arcmin$. The result is shown in Figure \ref{fig:power_spectrum}, along with an inset panel showing the $BB/EE$ ratio as a function of multipole $\ell$. In both $EE$ and $BB$, the model compensates the beam suppression of the input maps, recovering a power law out to $\ell\sim400$, corresponding to the $4\times$ the input $80\arcmin$ resolution, where the low-resolution spectra fall off exponentially. The $BB/EE$ ratio of the dust ViT map track closely that of the input GNILC map across multipole values, starting at the Planck measured value $BB/EE\sim0.65$ up to $\ell\sim200$, but drift off to unity at large $\ell$. This behavior is explained by the fact that our model is trained to undo beam smoothing suppression, which acts identically on $E$ and $B$ modes. As a result our model inherits the $BB/EE$ ratio of the GNILC maps rather than imposing one, whereas the PySM d10 small scales are injected with a fixed sub-unity $BB/EE$ ratio by construction. The instrumental noise and component separation residuals of the input GNILC map, which are statistically isotropic and contribute equally to $E$ and $B$, are likely retained in the dust ViT predictions and contribute to the rise of the $BB/EE$ ratio to unity at smaller scales. We discuss this limitation and potential future improvements in more detail in Section \ref{subsec:limitation}.

\begin{figure*}[t]
    \centering
    \includegraphics[width=0.98\textwidth]{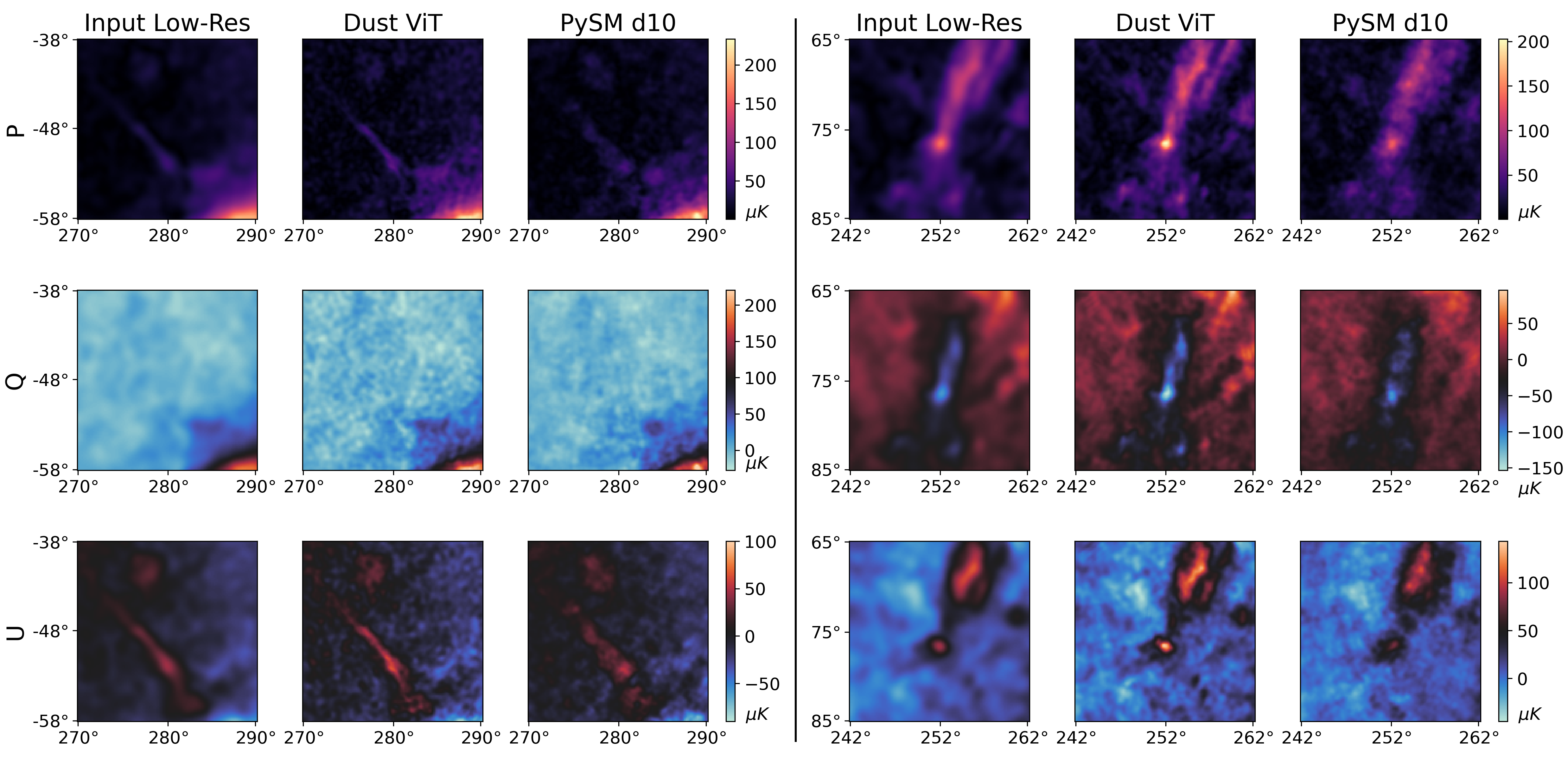}
    \caption{Comparison of Stokes $Q$, $U$, and polarized intensity $P=\sqrt{Q^2+U^2}$ for two sample $20\degree\times20\degree$ high-latitude regions. For each region from left to right: the low-resolution GNILC input at $80\arcmin$, the Dust ViT high-resolution prediction at $20\arcmin$, the PySM d10 model smoothed to $20\arcmin$. In regions with clear anisotropic features, the Dust ViT prediction reproduces coherent filamentary structures, whereas the PySM d10 model exhibits more disconnected, diffuse features.}
    \label{fig:compare_pysm_patch}
\end{figure*}

\begin{figure*}[t]
    \centering
    \includegraphics[width=0.95\textwidth]{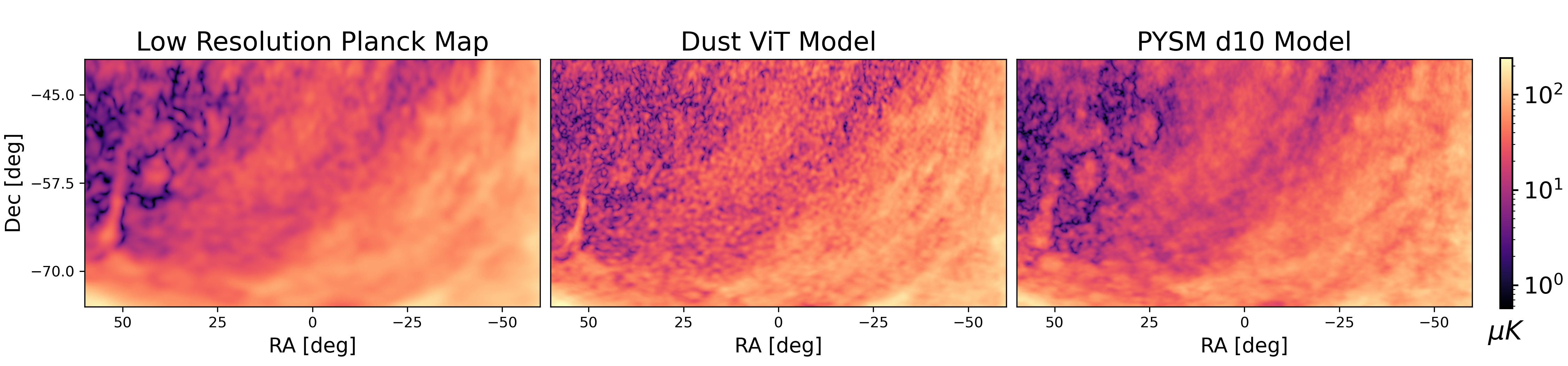}
    \caption{Polarized intensity map comparison in the BICEP/Keck field shown in log-scale. From left to right: the low-resolution GNILC Planck map at $80\arcmin$, the Dust ViT high-resolution prediction at $20\arcmin$, and the PySM d10 model smoothed to $20\arcmin$.}
    \label{fig:compare_bk_patch}
\end{figure*}

\begin{figure*}[t]
    \centering
    \includegraphics[width=0.9\textwidth]{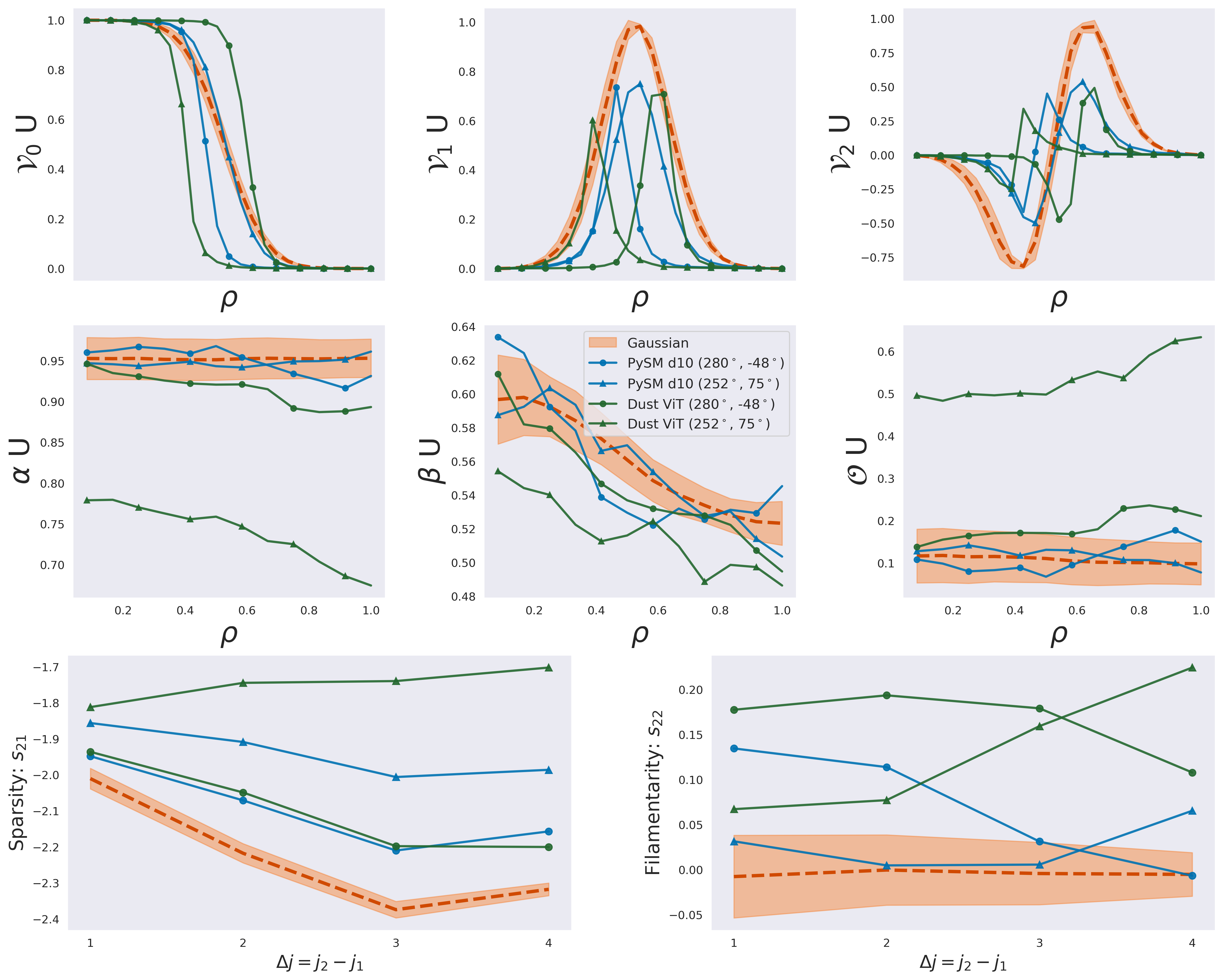}
    \caption{Scalar MFs, tensorial MFs, and ST statistics comparison for Gaussian, PySM d10, and dust ViT Stokes $U$ maps over the same $20\degree\times20\degree$ regions in Figure \ref{fig:compare_pysm_patch}. \emph{Top row}: Scalar MFs $\mathcal{V}_0$, $\mathcal{V}_1$, $\mathcal{V}_2$ as a function of threshold $\rho$. \emph{Middle row}: the tensor MF alignment $\alpha$, shape $\beta$, and orientation $\mathcal{O}$ parameters as a function of threshold $\rho$. \emph{Bottom row}: the ST sparsity $s_{21}$ and filamentarity $s_{22}$ as a function of scale separation $\Delta j=j_2-j_1$. In each case the orange dashed line and shaded band show the mean and $1\sigma$ of the Gaussian realizations. The solid blue and green correspond to the PySM d10 and dust ViT patches respectively. }
    \label{fig:mf_st_compare}
\end{figure*}

\section{Validation of High-Resolution Foreground Prediction} \label{sec:validation}

\subsection{Comparison with PySM and Gaussian Realizations} \label{subsec:pysm_comapre}

In Figure \ref{fig:compare_pysm_patch}, we compare the Stokes $Q$, $U$, and polarized intensity maps $P=\sqrt{Q^2+U^2}$ between the low resolution GNILC, the high resolution prediction from our dust ViT model, and the PySM d10 dust model \citep{borrill25-pm}, for two $20\degree\times20\degree$ sample regions in the northern and southern hemispheres at high latitudes. The PySM d10 model patches are smoothed to 20\arcmin\ to match the resolution of our maps. In Figure \ref{fig:compare_bk_patch}, we also show the native GNILC, PySM d10 and dust $P$ map comparison in log-scale in the BICEP/Keck region \citep{bicep22-ak}. the $\sim 400$ square-degrees patch of high latitude sky targeted by the BICEP and Keck Array CMB experiments. In regions of the input low resolution map where there are clear anisotropic features, the high resolution predictions of our dust ViT model show more coherent filamentary structures consistent with their low resolution counterparts, while the PySM d10 images at the same resolution exhibit more disconnected and diffuse features.

As \citet{borrill25-pm} emphasize, the PySM d10 model generates stochastic realizations of the small scales conditioned on a global power-law fit, and the combination of the log-polarization tensor formalism with amplitude modulation makes those realizations non-Gaussian and ties their amplitude to the large-scale template, but not their morphology to the actual sky. Our model instead conditions on multiple high-resolution tracers---with the \ion{H}{1}-based templates carrying most of the filamentary information (Figure \ref{fig:attention_image})---and is trained to undo beam smoothing and recover native GNILC maps from their smoothed counterparts, so that it solves a deconvolution problem rather than synthesizing small-scale features unconstrained by the data. As a result, the data-driven approach produces more connected anisotropic features at small scales, as observed in Figures \ref{fig:compare_pysm_patch} and \ref{fig:compare_bk_patch}. 

\subsection{Non-Gaussianity Comparisons on Small Regions} \label{subsec:mf_flat}

To quantify the observation that the dust ViT model produces more coherent, anisotropic features than the PySM d10 small-scale predictions, we apply MFs (scalar and MF) and ST statistics, to the sample patches in Figure \ref{fig:compare_pysm_patch}, comparing a Gaussian realization, PySM d10, and dust ViT maps. To isolate small-scale features, the scalar and tensorial MFs are computed on the high-pass filtered version of the maps. We adopt a similar procedure to \citet{borrill25-pm}, excluding multipoles $\ell<135$ in the maps using a sigmoid filter. The multipole threshold is selected to match the resolution of the input GNILC uniform resolution map at $80\arcmin$. In contrast, the ST statistics are computed on the full maps since they directly characterize structures at different scales. We show results for the $U$ channel maps, where Figure \ref{fig:mf_st_compare} shows contain the most prominent anisotropic structures for both regions. The $P$ and $Q$ channels show qualitatively similar results. The orange bands show the distribution over multiple random Gaussian realizations for these statistics, contrasted with those of the PySM d10 (blue) and dust ViT maps (green) in the sample $20\times20$ regions shown in Figure \ref{fig:compare_pysm_patch}. The MFs show a stronger departure from the Gaussian distribution for the dust ViT patches than the PySM patches, with the tensorial MFs specifically showing that the dust ViT is more anisotropic (lower $\alpha$ and $\beta$), with stronger preferred orientations (higher $\mathcal{O}$). The ST filamentarity statistic $s_{22}$ tapers off to zero at larger relative scales $\Delta j$ for the PySM d10 patches, but not for the dust ViT patches. As described in Section \ref{subsubsec:st}, larger $s_{22}$ characterizes stronger alignment of structures of a smaller scale $j_1$ onto a larger scale $j_2$. Thus a field dominated by thin filamentary structures will have high $s_{22}$ even at large $\Delta j=j_2-j_1$ scales comparable to the characteristic aspect ratio of the filaments. All together, the MFs and ST statistics show a consistent picture that both PySM and dust ViT are measurably more sparse than Gaussian random fields. However, the dust ViT model predicts more coherent, anisotropic and filamentary structures than the PySM d10 maps.

\begin{figure*}[t]
    \centering
    \includegraphics[width=0.9\textwidth]{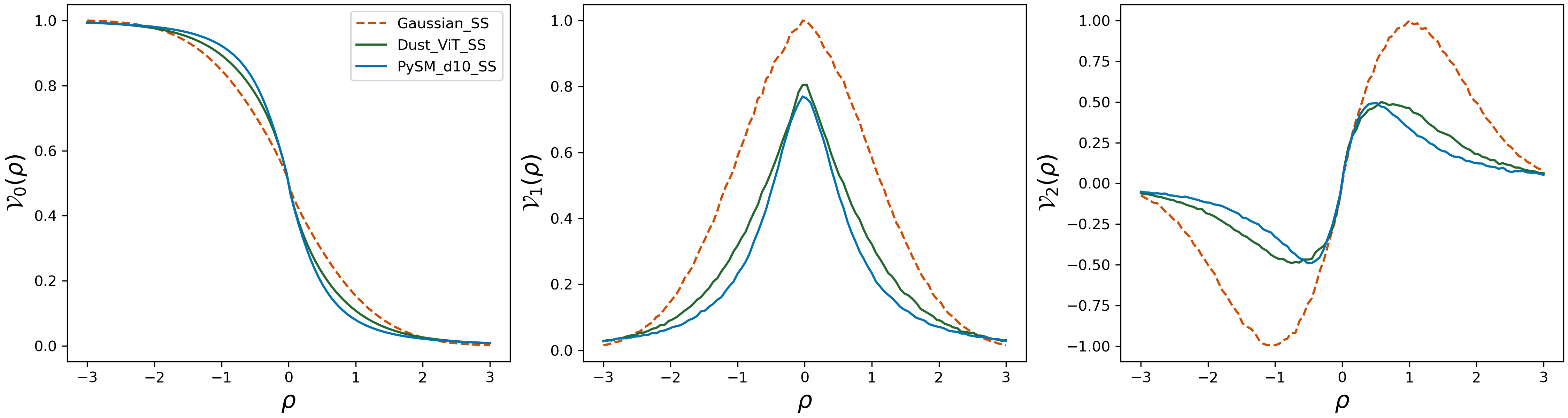}
    \caption{Scalar MFs as a function of threshold $\rho$ for the small-scales of the Gaussian, PySM d10, and dust ViT $Q$ maps on the sphere with a $|b|>30\degree$ mask. The small scale maps are created by excluding multipoles $\ell<135$ in the maps, $\sim$ the resolution of the input GNILC map to the dust ViT model. MFs for all three maps are normalized by the maximum of the Gaussian map across thresholds. }
    \label{fig:scalar_mf_sphere}
\end{figure*}

\begin{figure*}[t]
    \centering
    \includegraphics[width=0.9\textwidth]{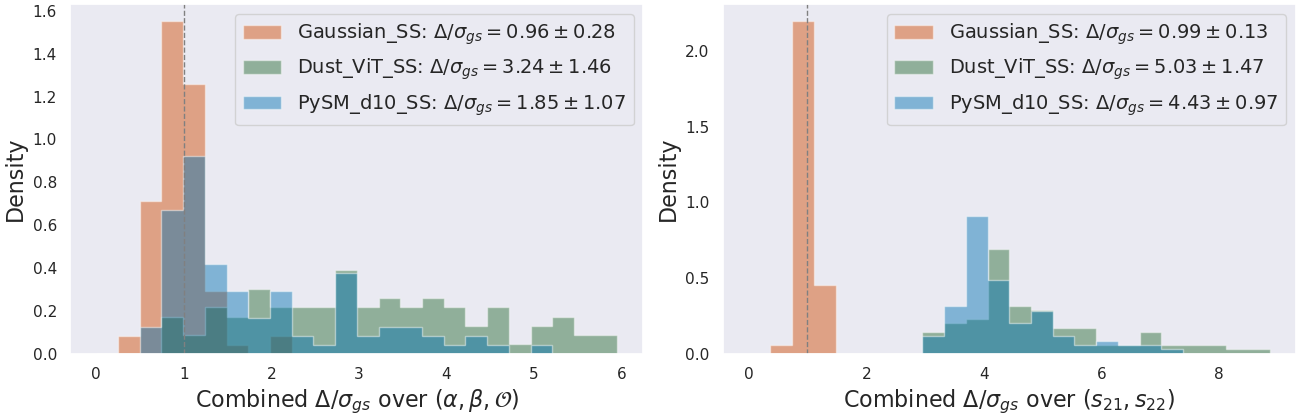}
    \caption{Distribution of the combined measure of the deviation of tensorial MFs (left) and ST statistics (right) of the dust ViT, PySM d10, and Gaussian random field patches from that of the Gaussian mean, measured across 96 $20\degree\times20\degree$ patches that cover the high-latitude ($|b|>30\degree$) sky. While both PySM d10 and dust ViT maps are significantly non-Gaussian, the dust ViT patches deviate more from the Gaussian random field baseline and has a larger variance across high latitude patches. }
    \label{fig:non_gauss_summary}
\end{figure*}

\subsection{Non-Gaussianity Comparison over High-Latitude Sky} \label{subsec:mf_sphere}

The example patches in Figure \ref{fig:compare_pysm_patch} are specifically smaller regions where the intensity is dominated by strongly anisotropic features. In those cases, the small-scale predictions of PySM d10 and dust ViT both deviate significantly from the Gaussian distributions, but the dust ViT maps exhibit more coherent anisotropic structures. \citet{borrill25-pm} found that while the PySM d10 model deviates significantly from both a pure Gaussian realization and large-scale amplitude modulated Gaussian realization in specific small $20\degree\times20\degree$ regions, when examining large areas of the high latitude sky, the MF distributions between PySM and modulated Gaussian are much more consistent. This indicates that the non-Gaussianity in the PySM d10 map across large areas is dominated by amplitude modulation rather than the predicted small-scale structures. To compare our model over a large area, We compute the scalar MFs of the three Q maps directly on the sphere over the full high latitude sky with a $|b|>30\degree$ mask using the \texttt{pynkowski} algorithm  \citep{carones24-pynkowski}. Figure \ref{fig:scalar_mf_sphere} shows that over large areas, which includes many regions with much lower SNR than the Figure \ref{fig:compare_pysm_patch} patches, the scalar MFs of the PySM and dust ViT maps both deviate significantly from pure Gaussian realization but are much more comparable with each other. However, in contrast to PySM, which imposes a large-scale amplitude modulation directly to the generated small-scale structure, the dust ViT model learns its small-scale non-Gaussianity from training across image patches at different resolutions using information from different input tracers. 

A further caveat to this comparison is that PySM d10 small scales are drawn from a power law fitted over $\ell=20-69$, where the GNILC maps are signal dominated, and extrapolated to higher multipoles. As a result the observed high-$\ell$ field is extrapolated rather than reproduced, so the synthetic small scales are more noise free by construction. In contrast, the dust ViT is trained to undo a beam, using a set of tracers with their own systematics. The resulting small-scale predictions will retain noise and component-separation residuals of the input GNILC map, which while negligible for signal-dominated regions shown in Figure \ref{fig:compare_pysm_patch}, can reduce the non-Gaussianity of the dust ViT maps and bias the comparison with a more noise-free PySM map when evaluated over the full high-latitude sky. We discuss this limitation and potential future improvement in more detail in Section \ref{subsec:limitation}.

For tensorial MFs and the ST statistics, in the absence of an efficient algorithm to compute them over large areas on the sphere, we compute these coefficients over partially overlapping $20\degree\times20\degree$ patches over the full high-latitude sky, and derive from them aggregate measures of non-Gaussianity. The patch centers are created following the same procedure for creating the training patches described in Section \ref{subsec:pre-process}, with neighboring patches overlapping by 30\%. We experimented with different patch overlap and found similar results. This results in a total of 96 patches across the full sky region above $|b|>30\degree$. From the coefficients for dust ViT, PySM, and Gaussian random field patches, we compute $\Delta/\sigma_{gs}$ where $\Delta$ is the difference between the coefficients of individual ViT/PySM/Gaussian patches and the Gaussian realization mean, and $\sigma_{gs}$ is the standard deviation over the Gaussian realization patches. One combined $\Delta/\sigma_{gs}$ each is derived for the tensorial MF and ST statistics respectively by taking the root mean square (RMS) over individual coefficients. By construction, the Gaussian random fields have $\Delta/\sigma_{gs}\sim1$. In contrast, the PySM d10 patches have an RMS $\Delta/\sigma_{gs}$ of $1.85\pm1.07$ and $4.43\pm0.97$ for tensorial MF and ST statistics respectively, while the dust ViT patches have RMS $\Delta/\sigma_{gs}$ of $3.24\pm1.46$ and $5.03\pm1.47$ . The reported numbers are the mean $\pm$ std of the $\Delta/\sigma_{gs}$ over all patches, with the full distribution shown in Figure \ref{fig:non_gauss_summary}. The distributions show that both PySM d10 and dust ViT are both significantly non-Gaussian as measured by tensorial MFs and ST coefficients. Dust ViT contains patches that deviate more significantly from the Gaussian random field baseline, but also larger variance over the high latitude sky.  These results are consistent with our result that dust ViT maps produce more coherently filamentary structures in more signal dominated, highly anisotropic regions, but the level of non-Gaussianity becomes more comparable to PySM d10 when averaged across large areas of the high latitude sky.

\section{Discussion} \label{sec:discussion}

Accurate modeling of polarized dust emission at high angular resolution is crucial for the next generation of CMB polarization experiments. Component separation pipelines are validated against simulated skies, and any mismatch between assumed and true small-scale foreground properties propagates directly into residual foreground contamination in the recovered CMB maps. The same small scales set the fidelity of lensing reconstruction and delensing, where dust mis-modeled as a cosmological signal can contribute significant bias. In this work, we present a novel data-driven foreground modeling approach that fuses together multiple high-resolution dust polarization tracers to predict realistic, non-Gaussian foreground simulations at high angular resolution. 

\subsection{Comparison with Existing Small-Scale Models} \label{subsec:compare_models}


Alternative approaches in existing literature include the application of generative models such as generative adversarial networks (GANs) \citep{goodfellow14-gan}. ForSE \citep{krachmalnicoff21} trains a GAN to generate $12\arcmin$ structure from $80\arcmin$ Planck GNILC maps, calibrating the non-Gaussianity of its output against the Minkowski functionals of the \emph{total intensity} maps, and ForSE+ \citep{yao24forse} iterates the same procedure to reach $3\arcmin$ under an assumption of scale invariance. These models reach finer scales than we consider here, but they inherit two structural limitations. First, the conditioning remains a global summary statistic, so as with PySM d10 the generated features are constrained in their statistics rather than their morphology. Second, the statistics are measured in total intensity and then imposed on $Q$ and $U$. Total intensity is a positive-definite scalar that accumulates monotonically along the line of sight, whereas polarized emission is a spin-2 quantity whose contributions from regions with differing magnetic field orientations partially cancel. Polarized emission is consequently sensitive to the three-dimensional tangling of the field in a way that total intensity is not \citep{clark19-hs, cuckierman23-hs, halal24-hi}, and as a result there is strong reason for the two quantities to differ in their level of non-Gaussianity. 

In contrast, instead of the global statistics-based approaches adopted so far, our dust ViT approach is entirely data-driven and recovers coherent structures. Rather than synthesizing small scales consistent with a target statistic, the network is trained to recover a native-resolution GNILC map from a smoothed version of itself, conditioned on tracers that are themselves observed at high resolution. The task is a deconvolution rather than a generation problem, and the predicted structures are therefore coherent with the large-scale features they extend by construction, as the sample regions in Figure \ref{fig:compare_pysm_patch} illustrate. The attention maps in Figure \ref{fig:attention_image} additionally offer a degree of interpretability that generative models do not, making explicit which tracer supplies the information behind a given prediction and confirming that the \ion{H}{1}-based templates carry much of the filamentary content.

\subsection{Limitations and Future Outlook} \label{subsec:limitation}

Our current model is deterministic and produces a single realization for a given input, whereas the intended application to component separation requires an ensemble over which foreground uncertainty can be marginalized. A natural extension is to replace the output head with one that predicts a per-pixel mean and variance and to train under a Gaussian negative log-likelihood loss \citep{kendall17-uncertainty}, which would both supply calibrated uncertainties and identify the regions where the small-scale prediction is least constrained by the input tracers.

As discussed in Section \ref{subsec:mf_sphere}, the dust ViT model inherits the noise, systematics, and component-separation residuals of the GNILC maps it is trained to reproduce. An extension to the current training processes described in Section \ref{subsec:pre-process} that can address this issue is to use the Planck half-mission or detector-set splits: training with one split as input and the other as target leaves the sky signal common to both while the noise realizations are independent, so a model trained under a pixel-wise loss converges toward the noise-free mean rather than reproducing the noise. 

Moreover, our model operates at a single frequency, predicting high-resolution 353\,GHz maps from 353\,GHz inputs. One of the most consequential properties of dust emission for CMB analyses is that its spectral behavior varies across the sky, so that maps at different frequencies are not perfectly correlated \citep{pelgrims21-ls, mcbride23-ls}. Scaling our maps to other frequencies with a modified blackbody is straightforward, but such a scaling is spatially coherent by construction and therefore produces no decorrelation, limiting the use of the resulting simulations for testing the biases that frequency decorrelation induces. Extending the framework to predict several frequencies jointly, with the spectral parameters treated as additional output channels, is a natural future direction.

Finally, a direct followup work is to push the predictions to higher angular resolution. Currently, the output resolution is limited by the resolution of the input \ion{H}{1} maps at 16\arcmin. Future work can utilize \ion{H}{1}-based polarization maps built from the higher-resolution GALFA-\ion{H}{1} at 4\arcmin\ in the covered northern sky footprint \citep{Peek2017-ql}, and explore generalization to other regions with the lower-resolution \ion{H}{1}4PI-based map assuming partial scale invariance, supported by the Figure \ref{fig:attention_averaged} result that input resolution/scale does not affect the prediction of high resolution maps. 

\section{Conclusions} \label{sec:conclusion}

In this work, we propose a data-driven super-resolution framework for polarized thermal dust emission that combines multiple high-resolution tracers of the diffuse ISM, and demonstrate that it can predict small-scale dust polarization structure that is morphologically coherent with the large-scale features it extends. We apply the method to produce high-resolution $Q$ and $U$ maps at 353\,GHz over the high-latitude sky, and validate them against Gaussian realizations and the latest PySM dust model using Minkowski functionals and scattering transform statistics. Here is a summary of our main results:

\begin{itemize}
    \item We introduce a transformer-based model that fuses five input channels ---the Planck optical depth $\tau_{353}$, the \ion{H}{1}-based Stokes templates $Q_{HI}$ and $U_{HI}$, and the coarse GNILC $Q_{353}^{LR}$ and $U_{353}^{LR}$---together with an explicit target-resolution embedding, and predicts $Q_{353}^{HR}$ and $U_{353}^{HR}$ at $4\times$ the resolution of the input Stokes maps. The model is trained to recover native-resolution GNILC maps from smoothed versions of themselves under a pixel-wise $L_1$ loss. Unlike previous approaches that synthesize small scales to match a prescribed summary statistic, our model solves a deconvolution problem, so the predicted structures are coherent with the observed large-scale features by construction rather than by calibration.

    \item We train and validate the model simultaneously across four input resolutions of 60\arcmin, 80\arcmin, 100\arcmin, and 120\arcmin, with targets at 15\arcmin, 20\arcmin, 25\arcmin, and 30\arcmin\ respectively. The predicted maps agree closely with the ground truth, with an $L_1$ validation loss of 0.003 for images normalized between $-1$ and $1$, and with no significant difference in performance across resolutions. The attention maps additionally provide a degree of interpretability uncommon in generative foreground models: the \ion{H}{1}-based templates carry much of the information behind filamentary structure predictions, while the negligible weight assigned to the target-resolution embedding indicates that the high-resolution mapping is approximately scale invariant.

    \item We apply the model to the uniform-resolution GNILC maps at 80\arcmin\ to produce continuous $Q$ and $U$ predictions at 20\arcmin\ over the high-latitude sky ($|b|>30\degree$). We quantify the non-Gaussianity of the predicted small scales using scalar and tensorial MFs and ST statistics, comparing against Gaussian realizations and the PySM d10 model over sample $20\degree\times20\degree$ high-latitude regions. The dust ViT maps depart from the Gaussian expectation more strongly than the PySM d10 maps, with the tensorial MFs indicating greater anisotropy, and the ST filamentarity statistic $s_{22}$ remains elevated out to large scale separations $\Delta j$ for the dust ViT maps, consistent with a field dominated by thin, coherent filaments.

    \item Computing the scalar MFs directly on the sphere over the full high-latitude footprint, we find that the dust ViT and d10 maps become much more consistent, indicating that the non-Gaussianity averaged over large areas is comparable between the two maps. The distinction is in how that non-Gaussianity arises: PySM d10 imposes it by construction through amplitude modulation, whereas our model is given no such prescription and instead recovers it from training across different resolution regions using information from input dust and \ion{H}{1} tracers. 

    \item These results demonstrate the potential of data-driven, learned deconvolution as a general approach to foreground modeling, in which small-scale structure is inferred from multiple high-resolution tracers rather than drawn from a prescribed statistical prior. The principal limitations of the current model are that it is deterministic and produces a single realization, that it inherits the noise and component-separation residuals of the GNILC maps it is trained to reproduce, and that it operates at a single frequency and therefore cannot by
    itself capture the frequency decorrelation of dust emission. Follow-up work will address these by predicting per-pixel uncertainties under a Gaussian negative log-likelihood, exploiting Planck half-mission splits to decouple the target noise from the sky signal, and extending the framework to predict multiple frequencies jointly.
\end{itemize}

We will make the dust ViT model discussed in this work, in addition to the high-resolution 353\,GHz polarization maps, publicly available upon publication.

\vspace*{5mm}

\begin{acknowledgments}
We are grateful to Doug Finkbeiner, Sihao Cheng, and Brandon Hensley for helpful conversation. This work was supported by the National Science Foundation under CAREER grant No. AST-2441452. S.E.C. acknowledges additional support from an Alfred P. Sloan Research Fellowship. M.L. acknowledges support from a 2MB Graduate Fellowship for Frontier Research.
\end{acknowledgments}

%

\software{
    HEALPix \citep{gorski05-hp},
    NumPy \citep{harris20-numpy},
    PyTorch \citep{paszke19-pytorch},
    PySM3 \citep{thorne17-pm, zonca21-pm},
    reproject \citep{robitaille20-reproject},
    SciPy \citep{virtanen20-scipy},
    scattering \citep{cheng2021-si},
    pynkowski \citep{carones24-pynkowski}
}


\begin{table*}[t]
\centering
\caption{Fiducial model hyperparameter setup \label{tab:hyper}}
\setlength{\tabcolsep}{6pt}
\begin{tabular}{@{}l@{\hspace{2em}}l@{}}
\hline\hline
Quantity & Value \\
\hline
Encoder/decoder channel width $C$ & 32 \\
Residual blocks per encoder & 4 \\
Residual blocks per decoder & 5 \\
Transformer layers $L$ & 1\\
Attention heads $h$ & 1\\
Learning rate $\eta$ & $10^{-4}$ \\
Number of Epochs & 1000 \\
AdamW $(\beta_1, \beta_2)$ & $(0.9,\,0.999)$ \\
Train / val / test fraction & $0.8\,/\,0.1\,/\,0.1$ \\
Fiducial loss & MAE on normalized $Q$, $U$ \\
\hline
\end{tabular}
\end{table*}

\appendix

\section{Model Implementation and Hyperparameter Settings} \label{appx:hyperparameter}
Here we provide additional implementation details for the model described in Section \ref{subsec:model}, following \citet{halal24-pd}. Table \ref{tab:hyper} summarizes the hyperparameter and optimization settings.

Normalization is performed over the selected sky footprint rather than independently for each patch. We use three normalization factors: one for $\tau_{353}$, one shared by $Q_{HI}$ and $U_{HI}$, and one shared by the low- and high-resolution GNILC Stokes maps. Each factor is determined from the 99.9995th percentile of the corresponding absolute-value distribution, with larger magnitudes clipped before scaling. All augmented versions of a parent patch are assigned to the same training, validation, or testing subset.

All convolutions use $3\times3$ kernels, unit stride, and one-pixel zero padding, preserving the input pixelization. GELU nonlinearities and skip residual blocks are used in the encoders and decoders. At each pixel, the five encoded tracer vectors and the target-resolution embedding form a sequence of six vectors. The transformer therefore applies attention across tracers at the same pixel, while neighboring spatial information is incorporated through the convolutional layers. Layer normalization precedes the attention and feed-forward operations, surrounded by residual connections. Training minimizes the mean absolute error over the normalized $Q$ and $U$ outputs, without explicit constraints on their power spectra or morphological statistics. 


\bibliography{dust_vit}{}
\bibliographystyle{aasjournalv7}



\end{document}